\documentclass[aps,prd,twocolumn]{revtex4}

\usepackage[english]{babel}
\usepackage[final]{graphicx}
\usepackage[backref,draft=false]{hyperref}

\begin{document}

\title{A fast apparent horizon algorithm}
\author{Erik Schnetter\footnote{Email: schnetter@uni-tuebingen.de}}
\affiliation{Theoretische Astrophysik, Universit\"at T\"ubingen, 72076
T\"ubingen, Germany.  Web:
\href{http://www.tat.physik.uni-tuebingen.de}{http://www.tat.physik.uni-tuebingen.de}}
\date{2002-05-12}

\begin{abstract}
I present a fast algorithm to find apparent horizons.  This algorithm
uses an explicit representation of the horizon surface, allowing for
arbitrary horizon resolutions and, in principle, shapes.  Novel in
this approach is that the tensor quantities describing the horizon
live directly on the horizon surface, yet are represented using
Cartesian coordinate components.  This eliminates coordinate
singularities, and leads to an efficient implementation.  The apparent
horizon equation is then solved as a nonlinear elliptic equation with
standard methods.  I explain in detail the coordinate systems used to
store and represent the tensor components of the intermediate
quantities, and describe the grid boundary conditions and the
treatment of the polar coordinate singularities.  Last I give as
examples apparent horizons for single and multiple black hole
configurations.
\end{abstract}

\maketitle

\section{Introduction}

Apparent horizons serve several purposes during the numerical
evolution of general relativistic spacetimes.  In vacuum, spacetime
has no other prominent features that can be determined locally in
time, such as the shock fronts found in hydrodynamics.  Event horizons
are global structures, and it is not possible to find them during a
time evolution, because the future of the spacetime is not yet known.
Yet each apparent horizon indicates, under reasonable assumptions,
that there is an event horizon present somewhere at or outside the
apparent horizon.  Furthermore, apparent horizons are under certain
circumstances also isolated horizons, making it then possible to
calculate their mass and spin.

Locating apparent horizons is also necessary when one wants to apply
excision boundary conditions in a numerical evolution, where the
apparent horizon has to be used to determine the location and shape of
the excised regions.  It is therefore important to have a fast and
robust apparent horizon finder that is closely integrated with the
evolution code.

\section{Method}

An apparent horizon is an outermost 2-surface within a spacelike
hypersurface (time slice) with zero expansion and spherical topology.
The condition that it be an outermost such surface is difficult to
verify in practice, and I will not require it in the following.  That
means that in each point of the horizon, the condition
\begin{equation}
\label{horizon-function}
H := \nabla_i s^i + K_{ij} (s^i s^j - g^{ij}) = 0
\end{equation}
has to hold (see e.g.\ \cite[chapter 4]{thornburg-thesis}, or
\cite{ah-finding}).  Here $s^i$ is the (spacelike) outward normal to
the horizon, and $g_{ij}$ and $K_{ij}$ are the three-metric and
extrinsic curvature of the time slice.  $H$ is the so-called
\emph{apparent horizon function}.

An apparent horizon surface can be represented explicitely by a
function $h(\theta,\phi)$ that specifies (this is an arbitrary choice)
the radius $r$ as a function of the spherical coordinates $\theta$ and
$\phi$.  This choice restricts the possible surfaces to those of
star-shaped regions about the origin, but this restriction does not
cause problems in practice; other choices are equally possible.  The
horizon consists out of those points $(r,\theta,\phi)$ of the time
slice that satisfy the condition $r = h(\theta,\phi)$.

The apparent horizon can also be defined implicitly through an
expression $F(x^i) = 0$, with the level set function $F$ defined e.g.\
as
\begin{equation}
\label{implicit-location}
F(r,\theta,\phi) = r - h(\theta,\phi) \quad \textrm{.}
\end{equation}
This representation leads to the equation
\begin{equation}
\label{spacelike-normal}
s_i = \frac{\nabla_i F}{\sqrt{g^{jk} \nabla_j F \nabla_k F}}
\end{equation}
for the outward normal.  The function $F$ and the spacelike normal
$s^i$ can be calculated using either spherical coordinates $(r,
\theta, \phi)$, or transformed to use Cartesian coordinates $(x, y,
z)$.  Such coordinate distinctions are actually rather important for
the numerical implementation of an algorithm, so that I do want to
treat them in some detail.

The condition $H=0$ is a nonlinear elliptic equation in $h$, as can be
seen by substituting the definitions of $s^i$ and $F$ into $H$.  With
the above coordinate choice, the domain of the equation
$H(\theta,\phi) = 0$ is $(\theta,\phi) \in (0, \pi) \times [0, 2\pi)$
with coordinate singularities at the poles; the range of the equation
is $h \equiv r \in (0, \infty)$.

As alternatives to treating the apparent horizon equation as an
elliptic equation, as I do, there are also other methods for solving
it to be found in the literature.  Commonly used methods are mean
curvature flow \cite{sfb-382-mean-curvature-flow} or level flow
methods
\cite{flow-finder}, or minimisation procedures
\cite{anninos98:_findin_appar_horiz_dynam_numer_spacet}.  The
advantages of flow methods are that they are in general more robust in
finding an horizon, and that they are in practice able to find the
\emph{outermost} apparent horizon, given a large sphere as initial
data.  Their disadvantage is that they lead to parabolic or degenerate
elliptic equations that are expensive to solve.  Minimisation
procedures can also lead to a more stable formulation than elliptic
formulations, but are usually slower.  I will restrict myself to the
solution of the elliptic equation.  As described below, I have also
tried a simple Jacobi solver that is conceptually identical to a flow
method, and it shows the expected robustness.

It is possible to change the nature of equation
(\ref{horizon-function}) by extending it into the three-dimensional
time slice.  This is e.g.\ done by changing from the unknown function
$h(\theta,\phi)$ to a level set $h_\lambda(\theta,\phi)$, where
$\lambda$ is the level set parameter
\cite{sfb-382-mean-curvature-flow}.  The advantage of such an
extension is that multiple horizons can be found at the same time, and
that no initial guess for the horizon location is needed.  The
disadvantage is that such an extension will lead to a slower
implementation because of the additional dimension.

\section{Coordinates}

The above choice to use spherical coordinates to parameterise the
horizon introduces a preferred coordinate system into the otherwise
coordinate-independent equations (\ref{horizon-function}),
(\ref{implicit-location}), and (\ref{spacelike-normal}).  Another
preferred coordinate system comes from the grid used in the time
evolution code for the whole spacetime.  Often, $g_{ij}$ and $K_{ij}$
are given in Cartesian components on a Cartesian grid.  This raises
the question as to which coordinate system to use in the numerical
implementation of the horizon finder.  The coordinate choice will of
course not influence the physical results, but it can make the
implementation more complicated if it requires interpolation, or less
accurate if it leads to coordinate singularities.  It can also make
the results more or less difficult to interpret, as tensor quantities
are numerically always expressed through their coordinate components.
Tensors in different coordinate systems cannot easily be compared,
even if they are given at the same grid points.

I chose to represent the quantities on the horizon using their
Cartesian tensor components, even when they are given on grid points
on the horizon surface.  This has the advantage that the tensors
$g_{ij}$ and $K_{ij}$ do not have to have their components
transformed, although they do need to be interpolated to the apparent
horizon location.  It also has a disadvantage, because the partial
derivatives that are calculated on the horizon are given only in
spherical components, and those have then to be explicitely
transformed into Cartesian components.

The overall advantage of using Cartesian tensor components is that
non-scalar quantities on the horizon can more easily be compared to
quantities given in the whole spacetime, and that the coordinate
singularities at the poles do not influence the representation of such
quantities.  The singularities will of course still influence the
calculation of quantities on the horizon, but the representation of
the results will not have coordinate singularities any more.

Let me use the indices $i$, $j$, $k$ for Cartesian components ($i,j,k
\in [x, y, z]$), $u$, $v$, $w$ for spherical components in 3D ($u,v,w \in
[r, \theta, \phi]$), and $a$, $b$, $c$ for spherical components on the
horizon ($a,b,c \in [\theta, \phi]$).  This convention distinguishes
clearly which quantities are actually defined on what manifold, and
also makes coordinate transformations explicit.

\section{Discretisation}

The discretisation scheme of the 3D quantities, i.e.\ those living in
the whole time slice, does not play a role when locating a horizon.
There only has to be a way to interpolate these quantities onto the
horizon surface.  These quantities are usually discretised on a
Cartesian grid.  It is also possible to use spherical coordinates, or
to use mesh refinement, without influencing the way in which the
apparent horizon function $H$ is evaluated.

I chose to discretise the 2D quantities, i.e.\ those quantities living
on the horizon, by using a polar $\theta$-$\phi$-grid with constant
grid spacings $d\theta$ and $d\phi$.  A constant grid spacing works
well when the horizon has a shape that is not too far from a sphere.
Experiments show that distorted (peanut-shaped) horizons still work
fine.

Polar coordinates create coordinate singularities at $\theta=0$ and
$\theta=\pi$, which I avoid by having the grid points staggered with
respect to the poles.  I find that putting scalar quantities on such a
grid generally works fine and requires no special treatment near the
poles.  On the other hand, naively putting tensor components in
spherical coordinates on such a grid is a bad idea, as they either
become zero or diverge at the poles.

\begin{figure}
\includegraphics[scale=0.9]{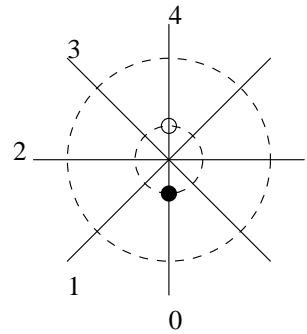}
\caption{\label{theta-boundary-1}North pole of the grid on the horizon,
in a polar projection.  This shows the grid lines as they lie on the
sphere.}
\end{figure}

\begin{figure}
\includegraphics[scale=0.9]{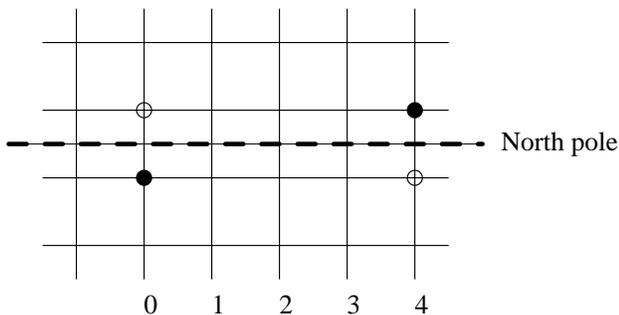}
\caption{\label{theta-boundary-2}North pole of the grid on the horizon,
as seen from the grid.  One can see that the black and white dots have
multiple images on the coordinate grid, separated by $\pi$ in the
$\phi$-direction.}
\end{figure}

The boundary condition in $\phi$-direction is periodicity: $f(\theta,
\phi) = f(\theta, \phi+2\pi)$ for any function $f$ living on the
horizon.  The boundary condition in $\theta$-direction, i.e.\ across
the poles, is a bit more involved.  It is $f(\theta, \phi) = P \cdot
f(-\theta, \phi+\pi)$ for arbitrary tensor components $f$, where the
parity $P = (-1)^r$ depends on the rank $r$ of the tensor of which $f$
is a component.  Figures \ref{theta-boundary-1} and
\ref{theta-boundary-2} demonstrate the polar boundary condition,
especially how the shift by $\pi$ in $\phi$-direction comes about.

This grid allows numerical partial derivatives to be taken only in the
$\theta$- and $\phi$-directions.  The apparent horizon equations
(\ref{horizon-function}), (\ref{implicit-location}), and
(\ref{spacelike-normal}) from above also need $r$-derivatives, and
those cannot be taken numerically.  However, the $r$-derivative of the
coordinate transformation operators are known analytically, and one
can see from the definition of $F$ that $\partial_r F = 1$ by
construction.  Finally, the partial derivative of the three-metric,
which is given in Cartesian coordinates, is calculated before the
metric is interpolated onto the horizon.

My choice of evaluating the apparent horizon function directly on the
grid that forms the horizon surface seems to be an uncommon one.  It
is also possible to evaluate the apparent horizon function not on the
surface itself, but instead in the three-dimensional spacetime on
those grid points that are close to the horizon \cite{ah-finding,
huq-finder, flow-finder}.  This method requires interpolating in both
directions between the surface and the time slice.  It has the
advantage that it is basically independent of the apparent horizon
discretisation, but it is also more expensive.  In addition, the
domain of the equation, i.e.\ the set of active grid points, changes
with the horizon surface, leading to further complications.

While I represent the apparent horizon shape using an explicit grid,
one different commonly used way is to expand the horizon surface
location in spherical harmonics \cite{baumgarte96:_implem,
gundlach98:_pseud, anninos98:_findin_appar_horiz_dynam_numer_spacet,
alcubierre00:_test}.  Compared to using a grid on the horizon surface,
an expansion in spherical harmonics has the advantage of having no
coordinate singularities.  Furthermore, one has direct insight into
and control over the high spatial frequency components.  Such an
expansion would lend itself naturally to a multigrid algorithm.  A
multipole representation is by principle restricted to star-shaped
regions, while an explicit representation is generic; however, this
disadvantage is not important in practice.  More problematic is that
integrations over the surface are rather expensive, as they scale with
$O(l_\mathrm{max}^4)$, were $l_\mathrm{max}$ is the number of
multipole moments used.  With an explicit grid, these integrations
scale with $O(n)$, where $n$ is the number of grid points, and where
one should choose $n = O(l_\mathrm{max}^2)$ for a fair comparison.

\section{Evaluating the apparent horizon function $H$}

The apparent horizon finder uses the algorithm described below to
evaluate the apparent horizon function $H$.  This description lists
the important intermediate quantities used in the apparent horizon
finder, and explains the order in which they are calculated.  This
method to evaluate $H$ is then subsequently used to find a solution
for $h$.

Although the quantities $F$, $s^i$, and $H$ are defined everywhere in
space, they only need to be evaluated on the horizon, i.e.\ on the
surface determined by $h$.  They still need to be defined in a
neighbourhood of the horizon, so that their $r$-derivatives can be
taken.  Their continuation off the horizon is chosen arbitrarily, and
is implicitly determined by the above choice of spherical coordinates.
Thus the finder calculates quantities on the two-dimensional horizon
surface only, although the quantities live in three dimensions.

It would be a bad idea to calculate the derivative of $s^i$
numerically, as $s^i$ is itself already calculated as a derivative.
Taking a derivative of a derivative leads to an effective stencil that
is larger, and hence slower, and that contains elements with a weight
of zero.  These zero weights lead to an odd--even decoupling of the
grid points, which in turn leads to large errors.  Instead, one uses
the chain rule, and calculates $\nabla_i s^i$ directly from second
derivatives of $F$.  This makes it necessary to calculate numerically
$\nabla_i \nabla_j F$ and $\nabla_i s^i$ along with $\nabla_i F$ and
$s^i$.

I describe explicitely which quantities are actually stored, and in
which coordinate system the tensor quantities are calculated.

\subsection*{Algorithm}

\begin{enumerate}

\item Give, as fixed background quantities, the three-metric
$g_{ij}(x^k)$ and the extrinsic curvature $K_{ij}(x^k)$ in the whole
time slice.  These quantities do not change while the horizon is
located.

\item Determine a 2-sphere $h(x^a)$ for which the apparent horizon
function is to be evaluated.  This is done by an elliptic solver as
described below, and is outside the scope of this algorithm.

\item Calculate the horizon grid point locations in Cartesian
coordinates as $x^i(x^a)$, using the definition of $h$.

\item Interpolate the three-metric, the extrinsic curvature, and the
partial derivative of the three-metric onto the horizon.  These
quantities can conceptually be viewed as representing $g_{ij}(x^u)$
and $K_{ij}(x^u)$, but are stored as $g_{ij}(x^a)$, $K_{ij}(x^a)$, and
$\partial_k g_{ij}(x^a)$.  (The partial derivative is needed later.)
Note that these tensors now live on the horizon, but have their
components still in Cartesian components, preventing coordinate
singularities near the poles.

\item Define the implicit horizon location $F(x^u) = F(r, \theta,
\phi) = r - h(\theta, \phi)$.  This quantity is not stored
explicitely.

\item Calculate an outward-directed covector $\nabla_v F(x^u)$ that
is orthogonal to the surface.  By definition, $\partial_r F = 1$, and
$\partial_a F = - \partial_a h$.  This is stored as $\partial_v
F(x^a)$ and $\partial_v \partial_w F(x^a)$.  (The second derivatives
are needed later).  The calculation needs the values $\partial_b
h(x^a)$ and $\partial_b \partial_c h(x^a)$, which are calculated on
the horizon surface.

\item Define the transformation operator $T_i^u$ that transforms from
spherical to Cartesian covariant coordinate components.  This operator
is given by $T_i^u$ = $\partial x^u / \partial x^i$, where the
$x^i(x^u)$ are given by the usual
\begin{eqnarray}
\nonumber x	& =	& r \sin(\theta) \cos(\phi)	\\
\nonumber y	& =	& r \sin(\theta) \sin(\phi)	\\
\nonumber z	& =	& r \cos(\theta)
\end{eqnarray}
Evaluate this operator and its derivative at $r=h$, and store it as
$T_i^u(x^a)$ and $\partial_j T_i^u(x^a)$.

\item Transform the components of the outward-directed covector into
spherical coordinates: $\nabla_i F(x^u) = T_i^v\, \nabla_v F(x^u)$.
Store $\nabla_i F(x^a)$ and $\nabla_i \nabla_j F(x^a)$.  The
calculation of the second derivative needs the derivative of the
transformation operator $T$, and also needs the partial derivative of
the three-metric for the connection coefficients.

\item Normalise the outward-directed covector $\nabla_i F(x^u)$ and
raise its index to $s^i(x^u)$, using the three-metric $g_{ij}$.
Calculate its derivative $\nabla_i s^i(x^u)$.  This needs the second
derivative of $F$.  Store $s^i(x^a)$ and $\nabla_i s^i(x^a)$.

\item Calculate the apparent horizon function $H = \nabla_i s^i +
K_{ij} (s^i s^j - g^{ij})$.  Store it as $H(x^a)$.

\end{enumerate}

For the above calculation I use a second-order uniform Cartesian
interpolator, and second-order centred differences to approximate the
partial derivatives.  Altogether, the apparent horizon function $H$
thus depends on the background spacetime as given by $g_{ij}$ and
$K_{ij}$, and depends on the horizon location $h$.

\section{Finding the horizon}

In order to solve a nonlinear equation, one needs initial data, which
is in this case an initial guess for the horizon location.  This
initial data selects which apparent horizon will be found, if there
are several present in the time slice.  During the evolution one can
easily use the location from the previous time slice as initial data;
this works very well in practice and is called \emph{apparent horizon
tracking}.  However, finding horizons in an unknown spacetime is more
difficult.  Depending on how e.g.\ the initial configuration for an
evolution run is constructed, one does not know whether, where, or of
what shape the horizons are.  A similar problem is encountered during
the evolution, as new apparent horizons may appear.  One does not know
their location and not even their time of appearance in advance.

Shoemaker et al.\ \cite{flow-finder} have implemented a mechanism in
their apparent horizon finder that automatically detects whether a two
black hole system has a common horizon, and if not, finds the two
separate horizons instead.  I did not use such a mechanism; if a
horizon does not exist, I had my solver abort.  I found it to be
sufficient in practice to set up the finder manually with different
initial guesses, corresponding to the different suspected horizons in
a time slice.

\subsection{Newton-like method}

I implemented a fast Newton-like method by calling the PETSc library
\cite{petsc-home-page, petsc-manual}.  This library provides several
efficient parallel solvers for nonlinear elliptic equations.  It uses
Newton-like methods to reduce the nonlinear problem to a linear one,
and then offers Krylov subspace methods to solve it.

Additionally, PETSc offers to calculate numerically the Jacobian that
is necessary for a Newton-like method.  I estimate that this is
probably about an order of magnitude slower than a hand-coded
Jacobian; it is therefore not recommended by the PETSc authors.  On
the other hand is it much easier to implement, and most importantly
also much easier to get correct.  Thornburg \cite{ah-finding} explains
the necessary and tedious steps of implementing an explicit Jacobian
in detail.  Huq et al.\ \cite{huq-finder} use numerical perturbations
instead, and their explicit calculation of those seems to complicate
their implementation significantly.  Fortunately, PETSc is fast enough
in my case, so I never tried a hand-coded Jacobian.  Tracking a
horizon with a reasonable resolution takes only about half a second on
a notebook with yesteryear's hardware.

\subsection{Flow-like method}

Unfortunately, the apparent horizon equation is rather nonlinear, and
the radius of convergence of Newton-like methods seems to be very
small.  Instead of a fast Newton-like method, one can apply a slow but
robust flow-like method, which can be viewed as essentially an
iterative Jacobi-type relaxation method.  Such a method has a larger
radius of convergence, and is in practice reliable in finding the
outermost apparent horizon.  For this method one generally uses a
large sphere as initial data.  The horizon shape then flows gradually
inwards, following the expansion of the horizon surface
\cite{flow-finder}.

With the above coordinate choice, the grid spacing near the poles is
very small, and this requires very small convergence factors of the
order of $10^{-3}$ or less for a reasonable resolution.  Finding a
horizon with a coarse resolution takes several minutes on the above
hardware.  In principle one could first flow-find a coarser
approximation to the horizon, and then Newton-find the horizon with a
higher resolution, but I have not tried this.

I do not think that a multigrid algorithm can be used to accelerate
flow-finding, because the coarse resolution is already so coarse that
no coarser grid could reasonably describe a spherical shape.  It might
be worthwhile to use a different discretisation scheme for the horizon
surface, so that the grid spacing does not tend to zero near the
poles, or to use spherical harmonics as mentioned above.

\section{Analysing the horizon}

Once the location of the horizon is known, one can calculate its area
$A$ by integrating over the surface, and from that the horizon radius
$R$ via the definition $A = 4\pi R^2$.  The irreducible mass is given
by $R/2$, which is different from the total mass $M$ in case the spin
$a$ is non-zero.

One can also calculate the lengths of the equatorial and two polar
circumferences (at $\phi=0$ and at $\phi=\pi/2$).  These
circumferences can give a hint as to the coordinate distortion of the
horizon.  Of course, these hints are only reliable if one knows in
advance about the symmetries of the horizon, which is only the case in
testing setups.

If one assumes that the horizon has a spin that is aligned with the
$z$ axis, then one can use the area and the equatorial circumference
to calculate the spin magnitude.  In the elliptic coordinates that one
prefers for Kerr black holes (see e.g.\ \cite[section
3.3]{livingreviews-cook}), a horizon with mass $M$ and spin $a$ is
located at the coordinate radius $r = M + \sqrt{M^2 - a^2}$.  It has
the area $A = 4\pi (r^2 + a^2)$ and the equatorial circumference $L =
2\pi (r^2 + a^2)/r$.  (I thank Badri Krishnan for pointing out the
latter to me.)  These equations can be solved for $a$ and $M$.

It turns out that the estimate for the total mass $M$ that is
calculated this way has a reasonable accuracy, while the estimate for
the spin $a$ is rather inaccurate when $a$ is small.  In this case,
$a^2$ can even become negative.  This is mainly caused by accumulation
of numerical errors.  An $O(10^{-2})$ error in $a^2$ shows naturally
up as an $O(10^{-1})$ error in $a$.  Of course, the values for $a$ and
$M$ are only correct when the spin is indeed aligned with the $z$
axis, and that will usually not be the case (nor will it be easily
verifiable) in a spacetime that is given numerically.  However, if the
horizon should be isolated, then the isolated horizon formalism
\cite{generic-ih, rotating-ih} can provide a way of reliably
calculating the spin and thus the total mass of the horizon.

\section{Examples}

I have implemented this apparent horizon finder algorithm within the
Cactus framework \cite{cactus-home-page}, which has recently gained
some popularity in the numerical relativity community.  This will
potentially make it easier for other people to use my implementation
of this algorithm not only as an external programme, but also directly
from within their code.

Below I give results from applying this finder to configurations with
one and multiple horizons.  First I use single black holes in
Kerr--Schild coordinates as test data to judge the accuracy and
quality of this method.  Then I apply the finder to superposed
Kerr--Schild data to demonstrate the behaviour of this method in the
presence of multiple black holes, and compare the horizon masses to
the corresponding ADM quantities.

\subsection{Single black hole}

I performed a convergence test with Kerr--Schild data.  (See e.g.\
\cite[section 3.3.1]{livingreviews-cook} for a definition of the
Kerr--Schild coordinates.)  Table \ref{convtest-single-1-results}
contains results from runs with a black hole of mass $M=1$ and spin
$a=0$.  The resolution of both the 3D time slice and the 2D grid on
the horizon increases downwards.  The last line contains the analytic
values.  $R/2$ is the irreducible horizon mass, and $M_\mathrm{ADM}$
is the ADM mass (see e.g.\ \cite[eqn.\ (7.6.22)]{adm-mass}) of the
whole spacetime, which I calculated numerically for reference.

\begin{table}
\begin{tabular}{l|ll|ll}
run	& $dx$		& $d\phi$	& $R/2$		& $M_\mathrm{ADM}$	\\\hline
a	& $1/4$		& $2\pi/36$	& 0.991111	& 1.00043		\\
b	& $1/8$		& $2\pi/72$	& 0.997693	& 1.00010		\\
c	& $1/16$	& $2\pi/144$	& 0.999429	& 1.00002		\\
	& $0$		& $0$		& 1.0		& 1.0			
\end{tabular}
\caption{\label{convtest-single-1-results}Horizon and ADM masses for a
black hole with $M=1$ and $a=0$.  Both the spacetime and the horizon
surface vary in resolution.  The last row gives the analytic values.}
\end{table}

\begin{table}
\begin{tabular}{l|ll}
runs	& $R/2$	& $M_\mathrm{ADM}$	\\\hline
a, b, c	& 3.79	& 4.43			\\
a, b	& 3.85	& 4.37			\\
b, c	& 4.04	& 4.18			
\end{tabular}
\caption{\label{convtest-single-1-cfacts}Convergence factors for the
results from table \ref{convtest-single-1-results}.  A value of 4
indicates perfect second order convergence.}
\end{table}

In general, I would consider the resolutions $1/4$, $1/8$, and $1/16$
to be coarse, reasonable, and fine, respectively.  Corresponding time
evolution simulations need to run in the year 2002 on a notebook, a
workstation, or a supercomputer.

Table \ref{convtest-single-1-cfacts} shows the convergence factors for
three-way and two-way convergence tests from the above runs.  The
convergence factors $f$ are calculated in the usual way via $f =
(C-M)/(M-F)$ or $f = (C-A)/(F-A)$, where $C$, $M$, $F$, and $A$
represent the coarse, medium, fine grid, and analytic values,
respectively.  A factor of 4 indicates perfect second-order
convergence.  The results indicate that a spatial resolution of
$dx=1/4$ is not sufficient to be in the convergence regime, and also
that the numerical ADM masses reported here might not be very
accurate.

In another convergence test, shown in table
\ref{convtest-single-2-results}, I kept the spatial resolution of the
time slice constant at $dx=1/8$, and varied only the resolution of the
apparent horizon grid.  This time the black hole had a mass $M=1$ and
a spin $a_z=1/2$, again in Kerr--Schild coordinates.  The apparent
horizon spin $a_z$ is estimated via its equatorial circumference.  The
last line of the table shows the analytic values.

\begin{table}
\begin{tabular}{l|l|lll}
run	& $d\phi$	& $R/2$		& $a_z$		& $M$		\\\hline
A	& $2\pi/36$	& 0.960886	& 0.496507	& 0.994655	\\
B	& $2\pi/72$	& 0.964675	& 0.498183	& 0.998537	\\
C	& $2\pi/144$	& 0.965594	& 0.498815	& 0.999511	\\
D	& $2\pi/288$	& 0.965819	& 0.498997	& 0.999754	\\
	& 0		& 0.965926	& 0.5		& 1.0		
\end{tabular}
\caption{\label{convtest-single-2-results}Horizon masses and spins and
ADM masses for a black hole with $M=1$ and $a_z=1/2$.  Only the
horizon surface resolution varies.  The last row gives the analytic
values.}
\end{table}

\begin{table}
\begin{tabular}{l|lll}
runs	& $R/2$	& $a_z$	& $M$	\\\hline
A, B, C	& 4.13	& 2.65	& 3.99	\\
B, C, D	& 4.08	& 3.47	& 4.01
\end{tabular}
\caption{\label{convtest-single-2-cfacts}Convergence factors for the
results from table \ref{convtest-single-2-results}.  A value of 4
indicates perfect second order convergence.}
\end{table}

The convergence factors for this test are shown in table
\ref{convtest-single-2-cfacts}.  As the spacetime is now fixed, and
differs slightly from the analytic solution due to the discretisation
errors, I omit the two-way convergence tests.  It is clearly visible
that the spin estimate does not converge to second order.  This is
unfortunate, but not of much consequence, as this method for the spin
calculation is not applicable in general anyway.

\begin{figure}
\includegraphics{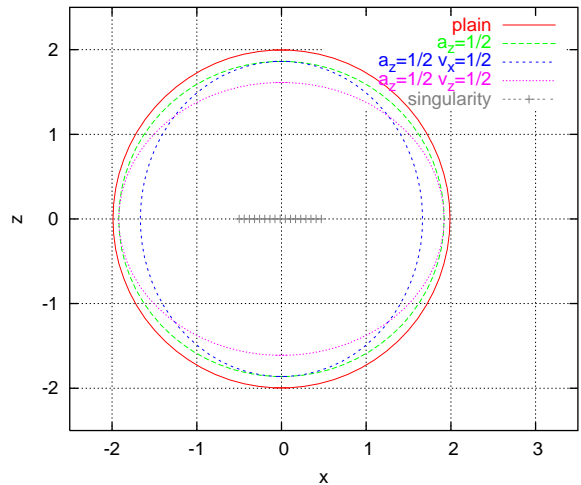}
\caption{\label{horizons}Apparent horizons for Kerr--Schild black holes
with varying spins and boosts.  Displayed is a cut through the $x$-$z$
plane.}
\end{figure}

Figure \ref{horizons} shows horizon shapes as reported by the finder.
It shows apparent horizons of four black holes with varying spins and
boosts.  The singularity is at the origin for $a=0$, and is a ring
with radius $a_z$ in the $x$-$y$ plane when the black hole is
spinning.  This figure shows nicely that excision boundary conditions
become more complicated for spinning black holes, as the apparent
horizon and the singularity can get rather close to each other.

\subsection{Multiple black holes}

In order to prepare and analyse initial data for binary black hole
collision runs, I superposed two Kerr--Schild black holes as proposed
in \cite{superposed-kerrschild}, and then solved the constraint
equations.  This results in spacetimes with either two separate or a
single distorted (``merged'') black holes.  One can freely specify the
locations, masses, and spins of the black holes prior to the
superposition.  The relation to the properties of the superposed black
holes is not known, but it is hoped the superposition will not change
them much.

For the superposition, I chose a grid spacing of $dx=1/4$, an outer
boundary that has a distance $8.0$ from the centres of the black
holes, and excised a spherical region with radius $1.0$ around the
singularities.  I superposed the three-metric and the extrinsic
curvature without attenuation.  I then solved the constraint equations
using the York-Lichnerowicz method with a conformal
transverse-traceless decomposition.  (This method is described e.g.\
in \cite{livingreviews-cook}, section 2.2.1.)  I used Dirichlet
boundary conditions for the vector potential and the conformal factor,
keeping the boundary values from the superposition.

Whether or not a common apparent horizon exists is a hint as to
whether the black hole configuration contains results from two
separate or a single merged black holes; this fact is not known
otherwise.  The only way to find out would be to evolve this
configuration for a long enough time so that the event horizon could
be tracked backwards in time.  This is unfortunately not easily
possible (if at all) with today's black hole evolution methods.

I created a series of initial data configurations for two equal-mass
($M=1$) non-spinning non-boosted black holes with varying distances.
Table \ref{horizon-masses} shows the common horizons masses $M_C =
R_C/2$, the individual horizons masses $M_I = R_I/2$, and the ADM
masses of the spacetimes $M_\mathrm{ADM}$.  The spins of the
superposed black holes are known to be zero for symmetry reasons.

\begin{table}
\begin{tabular}{l|lllll}
$z$	& $R_C/2$	& $R_I/2$	& $M_\mathrm{ADM}$	& $E_B$		& $E_R$		\\\hline
0.0	& 2.280		& ---		& 2.306			& ---		& +0.026	\\
1.0	& 2.245		& ?		& 1.973			& ?		& -0.272	\\
1.75	& 2.167		& 1.137		& 1.986			& -0.107	& -0.181	\\
2.0	& 2.130		& 1.087		& 1.989			& -0.044	& -0.141	\\
2.5	& 2.023		& 1.041		& 1.994			& -0.059	& -0.029	\\
3.0	& (2.101)	& 1.020		& 1.997			& (+0.060)	& (-0.104)	\\
4.0	& ---		& 1.002		& 2.000			& \multicolumn{2}{c}{-0.004}	\\
6.0	& ---		& 0.991		& 2.000			& \multicolumn{2}{c}{+0.018}	\\
8.0	& ---		& 0.989		& 2.000			& \multicolumn{2}{c}{+0.022}	
\end{tabular}

\caption{\label{horizon-masses}Masses of the common horizons, the
individual horizons, and the ADM masses for two superposed black holes
for varying distances.  $E_B$ is binding energy of the two black
holes, and $E_R$ is amount of radiation present in the spacetime, as
explained in the text.  For $z=\pm 3.0$, no common horizon was
detected any more.  The last three rows show the sum of $E_B$ and
$E_R$ only.}

\end{table}

The row $z=0.0$ contains two black holes that were superposed at the
same location, i.e.\ form a single black hole in a coordinate system
different from Kerr--Schild.  This is a consequence of the way the
extrinsic curvatures are superposed.  Note that the mass is not simply
twice the mass of a single black hole.  Also, as this spacetime is not
spherically symmetric due to the outer boundary condition, the ADM
mass and the horizon mass need not be the same.

In the row $z=\pm 3.0$, the apparent horizon finder did not converge,
but was close.  The $L_2$-norm of the residual was still $0.0248$ when
the solver aborted.  That means that from this row on there is no
common horizon any more, but the solution error here is still
comparable to the discretisation error.

The difference $E_B = M_C - 2 M_I$ can be interpreted as the binding
energy of the system.  (Here $M_C = R_C/2$ and $M_I = R_I/2$ because
the spin is zero.)  The binding energy magnitude seems to decrease
with distance, and is negative as it should be, except in the row
$z=\pm 3.0$ where there was no common horizon detected any more.

The difference $E_R = M_\mathrm{ADM} - M_C$ is the amount of radiation
present in the spacetime.  Of course, $E_R$ must be positive, meaning
that the negative values found numerically are unphysical and point to
numerical errors.  I suspect that the error lies mainly with the
calculation of the ADM mass.  I calculate the ADM mass as a surface
integral near the outer boundary.  When I evaluate the integral at
locations further inwards, closer to the apparent horizon, then the
ADM mass increases and gets closer to the horizon mass.  This is
inconsistent and cannot happen when the constraints are satisfied.
However, in a numerically given spacetime the constraints are only
satisfied up to the discretisation error.  It is thus in my opinion a
bad idea to calculate the ADM mass (which is a global quantity) from
information near the outer boundary only, relying on the constraints
to propagate the necessary information to the boundary.  A better
method to calculate the ADM mass is needed.  I decided to present the
ADM values here nonetheless, because they (although inaccurate) still
provide some insight into the spacetimes.

\begin{figure}
\includegraphics{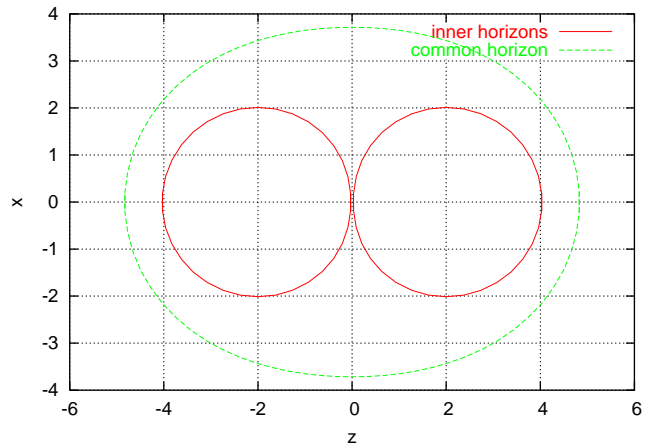}
\caption{\label{horizons2}Common and individual horizons for
superposed Kerr--Schild data at $z=\pm 2.0$.}
\end{figure}

\begin{figure}
\includegraphics{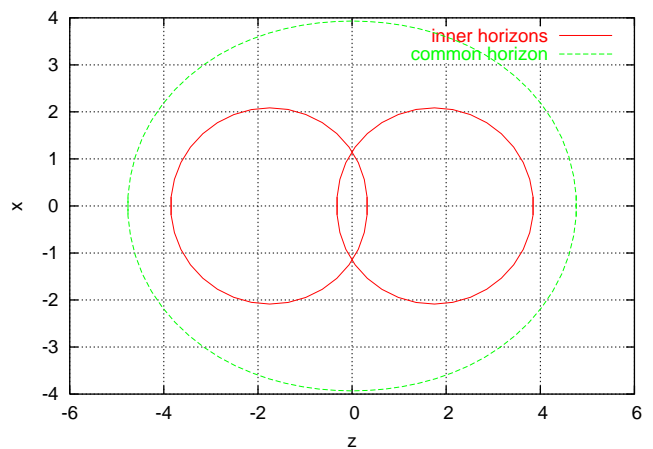}
\caption{\label{horizons2overlapping}Common and individual horizons
for superposed Kerr--Schild data at $z=\pm 1.75$.  The individual
horizons do overlap, and are surprisingly only minimally distorted.}
\end{figure}

Figure \ref{horizons2} shows the apparent horizons of two superposed
black holes with centres at $z=\pm 2.0$.  The common apparent horizon
indicates that there is a common event horizon present, and that this
time slice thus contains a single, merged black hole.  In figure
\ref{horizons2overlapping}, the black holes are closer together at
$z=\pm 1.75$.  The two inner apparent horizons actually do overlap;
this is not a numerical error.  It is surprising to see that the shape
of the horizons is not distorted more.  But unlike event horizons,
there is no reason why apparent horizons should deform or join when
two black holes approach each other.

\begin{figure}
\includegraphics{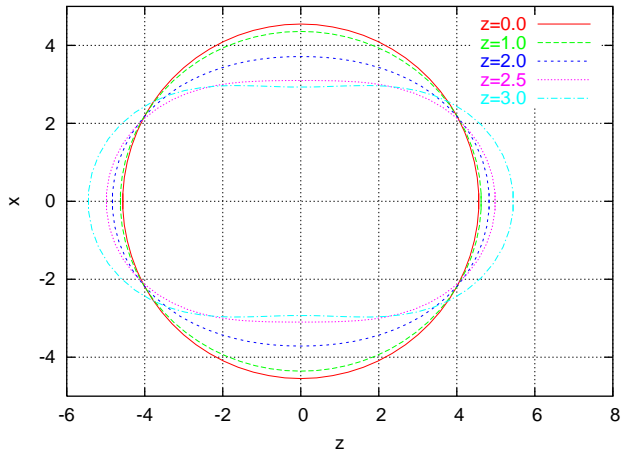}
\caption{\label{horizons2common}Common horizons for a series of merged
black holes with increasing distances.}
\end{figure}

\begin{figure}
\includegraphics{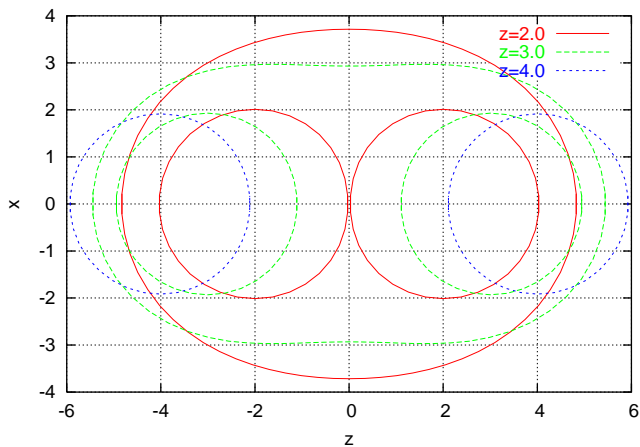}
\caption{\label{horizons2both}Individual and common horizons for
superposed Kerr--Schild black holes near $z=\pm 3.0$, where the common
apparent horizon disappears.}
\end{figure}

Figure \ref{horizons2common} shows the the common apparent horizons
for a series of black holes with increasing distances.  As the black
holes are placed further apart, the common horizons become
unsurprisingly more elongated.  Figure \ref{horizons2both} shows both
the individual and common black holes for varying distances for
comparison.  Near $z=\pm 3.0$, the common horizon ceases to exist.

\section{Summary}

Apparent horizons provide valuable insight into spacetimes, such as
mass and spin estimates for black holes.  They are indicators for
event horizons, and help keep track of the location of singularities.
They provide necessary information for the time evolution of black
hole spacetimes.  This apparent horizon finding algorithm provides a
fast way to track horizons during a numerical evolution, and also a
robust way to find horizons in initial data, or as they appear during
an evolution.

\section*{Acknowledgements}

I wish to thank to my collaborators Mijan Huq, Badri Krishnan, Pablo
Laguna, and Deirdre Shoemaker for countless inspiring discussions.
Hans-Peter Nollert provided helpful comments on this manuscript.
While working at this project, I was financed by the SFB 382 \glqq
Verfahren und Algorithmen zur Simulation physikalischer Prozesse auf
H\"ochstleistungsrechnern\grqq\ of the DFG.  Several visits to Penn
State University were supported by the NSF grants PHY-9800973 and
PHY-0114375.

\bibliographystyle{plain}
\bibliography{ah}

\end{document}